\documentclass[10pt, twocolumn]{amsart}

\usepackage{amsaddr}
\usepackage[authoryear,round]{natbib}
\usepackage{float}
\usepackage{hyperref}
\usepackage[pdftex]{graphicx}
\usepackage[a4paper, margin=2.0cm]{geometry}
\usepackage{bm}
\usepackage{graphicx}
\usepackage{booktabs}
\usepackage{caption}
\usepackage{orcidlink}
\DeclareMathOperator{\ex}{E}
\DeclareMathOperator{\var}{Var}

\begin{document}

\title{A Binomial-Like Probability Distribution With Heavy Tails}
\author{J\"{u}rgen Gro{\ss}{$^{\orcidlink{0000-0002-3861-4708}}$}}

\address{Institute of Mathematics, Mathematics Education \& Informatics Education University of Hildesheim, 31141 Hildesheim, Germany}
\email{juergen.gross@uni-hildesheim.de}


\date{}

\begin{abstract} A simple alternative to the binomial distribution that places more probability weight on the tails is considered. Its derivation only requires the weighted arithmetic mean of two discrete probability mass functions, one being the binomial itself and the other being the bi-uniform introduced here. Some properties are derived, and an application to a classical data set is discussed. The presentation  can be seen as an exemplary treatise on how to construct a statistical model, derive statistical properties, fit models to actual data by employing estimation methods, and verify appropriateness by using elements from statistical hypothesis testing.
\end{abstract}

\maketitle

\section{Introduction}

When discrete random variables and the probabilities of their possible outcomes  are introduced in an (advanced) undergraduate statistics course, the binomial distribution plays an important role. Whenever an authentic context resembles the ``flipping a coin'' situation with a never changing success probability $p$, then the number $X$ of successes from $n$ independent trials may be associated with the probability mass function (pmf)
$$
p_{X}(x) = P(X=x) = \binom{n}{x} p^{x} (1-p)^{n-x}
$$
for $x \in M=\{0,1,\ldots ,n\}$ and $p_{X}(x) = 0$ otherwise.
The expectation of $X$ is
\begin{eqnarray*}
\ex(X) & = & \sum_{x\in M} x \cdot p_{X}(x)\\
& = & \sum_{x=0}^{n} x \binom{n}{x} p^{x} (1-p)^{n-x} = n p\; ,
\end{eqnarray*}
where the last identity is often carried out as an example or exercise in a statistics course. Realizations of $X$ close to $np$ are most likely to appear, while values in the left and right tail, i.e. near $0$ and $n$, are endowed with rather small probabilities as long as $p$ is not quite close to either $0$ or $n$. The variance of the binomial distribution is
$$
v_{1}:= \var(X) = \ex(X^2) - [\ex(X))]^2 = n p (1-p)\; .
$$
See also Section 3.1 in \citet{hogg2019introduction} for derivations and further properties of the binomial distribution.

When there are $m$ observations $x_{1}, \ldots, x_{m}$ representing  the number of successes from $n$ trials, the binomial distribution is a natural candidate  to explain the data generating process. The method of moments estimate $\widehat{p}$ is obtained by solving the equation $\ex(X) = \overline{x}$, which gives
$$
\widehat{p} = \frac{\overline{x}}{n}, \quad \overline{x} = \frac{1}{m} \sum_{i=1}^{m} x_{i}\; .
$$
For the binomial model the method of moments estimator coincides with the estimator obtained by the maximum-likelihood method. The latter estimation principle will not be considered in this paper for simplicity.

It may very well happen that the data had not been generated by the pure ``flipping a coin'' random experiment, but by some related but slightly different process which may not entirely be transparent. The effect of a possible aberration is often noticed when observed and predicted frequencies are compared to each other, in which case the discrepancies  are tried to be explained from the data background. An often occurring situation is that of overdispersion, meaning that the data admits more variance than predicted from the binomial model. This usually manifests itself in the fact that although in principle observed and fitted distribution look quite similar, observed frequencies in the tails are larger than what were to be expected from the binomial model. Table \ref{tab:fisher} shows a classical data example of such a situation for the case $n=8$ and $m= 53680$ which is discussed in Section \ref{sec:app}.

\begin{figure}
	\includegraphics[width=\linewidth]{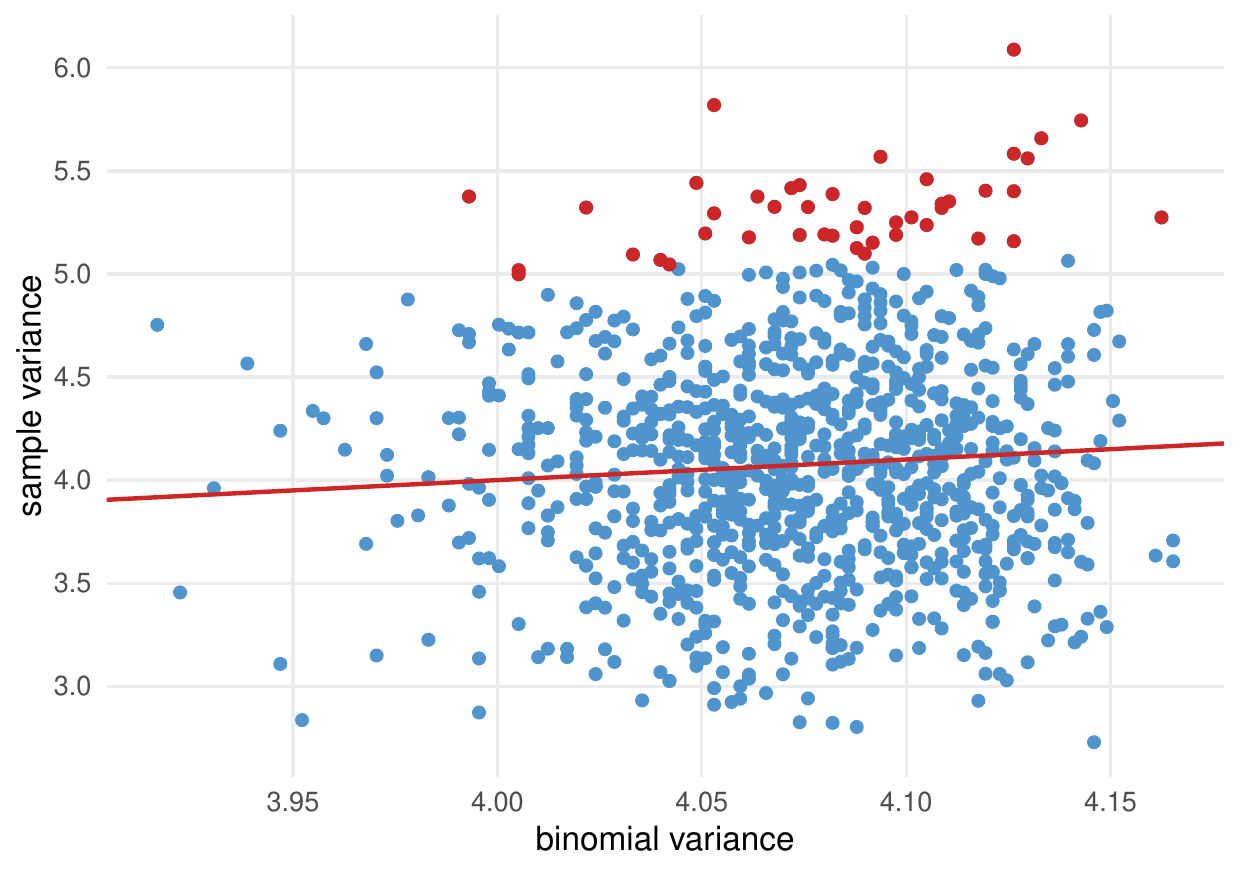}
	\caption{Variance estimates $\widehat{v}_{1}$ and $s^2$ for 1000 samples of size $m=100$ from a binomial distribution with parameters $n=17$ and $p=0.4$.
} \label{fig4} 
\end{figure}

\subsection{Overdispersion}\label{sec:over}

Before proceeding, some cautionary notes about the terms ``overdispersion'' and ``underdispersion'' seem to be in order. The most simple way to diagnose apparent overdispersion, also called extra-binomial variance, is to compare the estimated variance $\widehat{v}_{1}$, obtained by replacing $p$ with $\widehat{p}$ in $v_{1}$, with the sample variance $$
s^2 = \frac{1}{m-1} \sum_{i=1}^{m} (x_{i} - \overline{x})^2\; .
$$
It is important to note that when doing so one will almost always find either overdispersion or underdispersion, simply because the two variance estimators will not exactly be identical. 

For illustration purposes, this may be supported by a small simulation. For this, $m=100$ observations are 1000 times generated from a binomial distribution with parameters $n=17$ and $p=0.4$. Figure \ref{fig4} shows a scatter plot of the 1000 estimated variances  together with the angle bisector. There are 505 cases $s^2 > \widehat{v}_{1}$ above the bisecting line. None of these can indicate relevant overdispersion, as each sample stems in fact from the binomial distribution with variance $v_{1} = np(1-p) = 4.08$.

The ranges of the axes also imply that $s^2$ as an estimator has a much greater variance than $\widehat{v}_{1}$.

Of course, when the difference between $s^2$ and $\widehat{v}_{1}$ is only small, speaking of over- or underdispersion
is not sensible. Hence, some threshold is required on which reasonable decisions can be based. 

Statistical test procedures aim to deliver such critical values and one such procedure can be inferred from  \S 19, Sect. 19 in \citet{fisher1958statistical}. The binomial index of dispersion is defined as $$z = \frac{(m-1) s^2}{\widehat{v}_{1}}\; .$$
If it is larger than the
$(1-\gamma)$ quantile of the chi-square distribution with $m-1$ degrees of freedom for some specified significance level $\gamma$, this is an indication for extra-binomial variance.
For the simulated 1000 samples there are 45 samples exceeding the critical value when $\gamma=0.05$. This number is perfectly in accordance with the theory of statistical significance testing, allowing at most $\gamma \% $ incorrect rejections of the null hypothesis that the variance can be explained by the binomial distribution. The 45 cases are the red dots in Figure \ref{fig4}.

In his article on critical reading in statistics teaching, \citet{gelman2011going} discusses a textbook example in which a slight underdispersion is highlighted in an application of the binomial distribution to the sex ratio of births. \citeauthor{gelman2011going} rightly criticizes this as a misleading overemphasis of a circumstance which
turns out to be non-significant by the aforementioned test procedure. In light of the above one may add that even if there is a significant result, one should keep in mind that there still remains a small margin of error.

Section \ref{sec:app} also refers to a classical example on the sex ratio of births but from a different book and with respect to extra-binomial variance.

\subsection{Alternative Distributions}

An often proposed alternative when empirical frequencies do not resemble the  binomial probabilities is the so-called beta-binomial distribution. This distribution is briefly introduced in the following Section \ref{sec:betabin}.
Then, as the main objective of this paper, a further alternative is proposed. It is derived as a mixture (also sometimes called compounding) of the binomial with a quite simple second discrete distribution, which is introduced here and is called the bi-uniform distribution. Some properties of the latter are described in Section \ref{sec:biuni}. Section \ref{sec:mixture} is devoted to the mixture of the two, while  Section \ref{sec:app} shows an application to the classical data set mentioned above.

\begin{figure}
	\includegraphics[width=\linewidth]{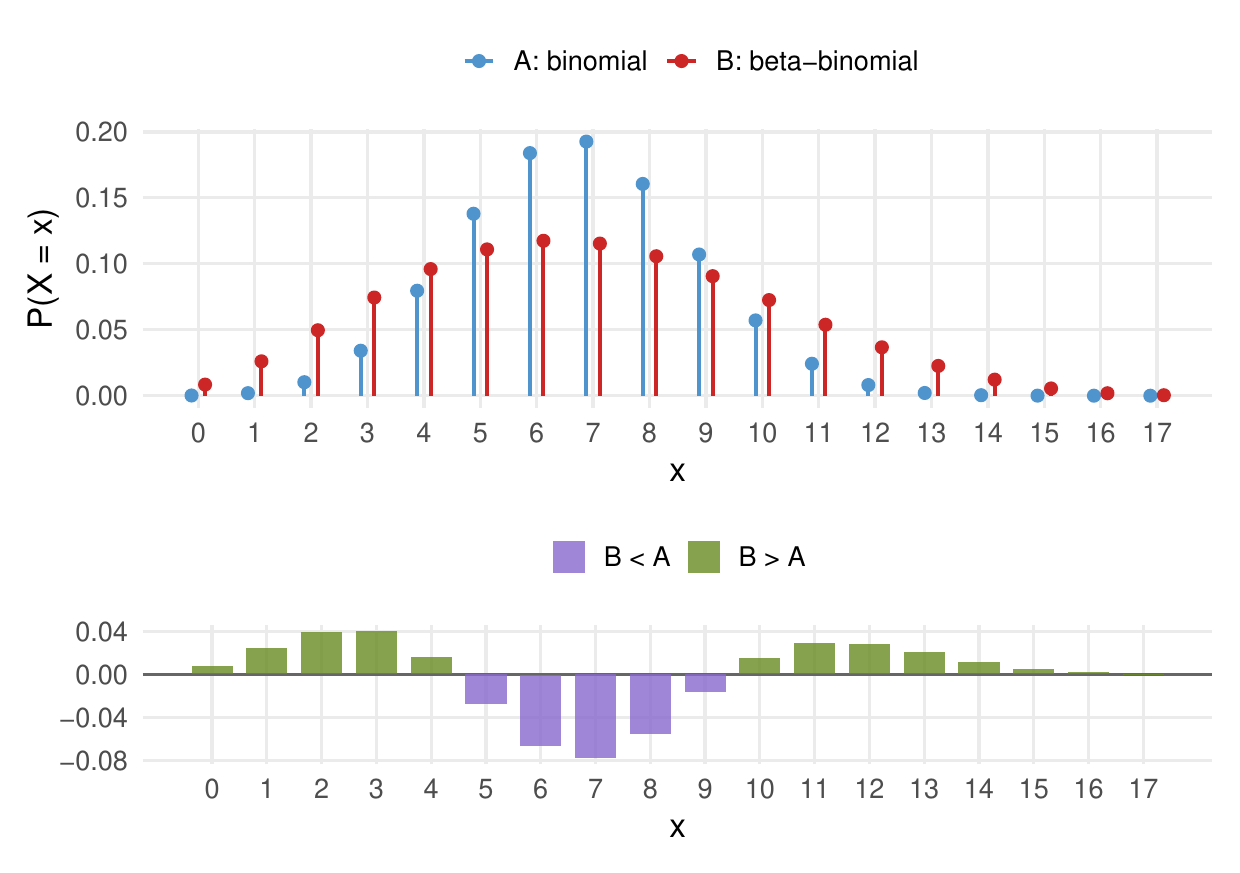}
	\caption{Graphical representation of probability mass functions for the case $n=17$ (upper panel) and their differences (lower panel).} 
	\label{fig3} 
\end{figure}

\section{Beta-Binomial Distribution}\label{sec:betabin}

When dealing with data for which the binomial distribution may be under consideration but extra-binomial variance is an issue, the most common recommendation is to apply the beta-binomial distribution. Its pmf is given as
$$
p_{X}(x) = \binom{n}{x} \frac{B(\alpha+ x,\beta + n-x)}{B(\alpha, \beta}
$$
for $x\in \{0,1,\ldots, n\}$ and $p_{X}(x) = 0$ otherwise. See \citet{skellam1948probability} for an often cited paper, although, according to \citet[Sect. 6.2.2]{johnson2005univariate}, there are earlier references containing similar derivations.

The parameters $\alpha$ and $\beta$ are positive real numbers and $$
B(\alpha, \beta) = \int_{0}^{1} t^{\alpha-1} (1-t)^{\beta-1} \,  d t
$$
is the so-called beta-function, see e.g. \citet{abramowitz1972handbook}. By letting
$$
\pi \equiv \pi(\alpha, \beta) := \frac{\alpha}{\alpha+ \beta}
$$
a beta-binomial distributed random variable $X$ has expectation $\ex(X) = n \pi$ and variance
$$
\var(X) = n \pi (1-\pi) \left(1 + \frac{n-1}{\alpha+\beta+1}\right)\; .
$$
The beta-binomial pmf may take various shapes. Figure \ref{fig3} shows a comparison with the binomial pmf for $n=17$ and $p=0.4$ for the choices $\alpha= 4$ and $\beta= 6$ so that $\pi(\alpha,\beta) = p$, implying that both pmfs have the same expectation. It is seen that the beta-binomial has less probability mass in the center while the tails are equipped with higher probabilities.

The beta-binomial pmf can be derived assuming that the parameter $p$ of the binomial pmf follows a continuous beta distribution with parameters $\alpha$ and $\beta$. Then the compounding  is achieved by integrating the binomial pdf weighted by the continuous beta probability density function with respect to $p$ over the interval from 0 to 1.
More explanations and details are given in Section 3.7 by \citet{hogg2019introduction}, where Example 3.7.3 specifically addresses the beta-binomial distribution. Since this procedure requires a more advanced understanding of mathematical statistics and calculus, it is usually not discussed in undergraduate statistics courses.

\subsection{Estimation}

Estimators of the parameters $\alpha$ and $\beta$ may be derived by the method of moments. This means that the estimators are obtained as solutions to the two equations resulting from setting expectation and variance equal to their
empirical counterparts $\overline{x}$ and $s^2$, respectively. There exist closed form solutions given as
$$
\widehat{\alpha} = \frac{(n - \overline{x} - s^2/\overline{x}) \overline{x}}{(s^2/\overline{x} + \overline{x}/n -1)n},
\quad
\widehat{\beta} = \widehat{\alpha} \, \frac{n -\overline{x}}{\overline{x}}
$$
see e.g. (6.86) in \citet{johnson2005univariate}. It is easily seen that
$$
\pi(\widehat{\alpha}, \widehat{\beta}) = \frac{\overline{x}}{n} = \widehat{p}\; ,
$$
implying that the estimated expectations of the fitted beta-binomial and binomial pmf will exactly be the same.

In the following, a mixture of the binomial with a second discrete pmf on the same support is proposed. It only uses one weight factor, so that the mixture simply becomes the weighted arithmetic mean of two probabilities for each of the points $x\in M=\{0,1, \ldots ,n\}$.
In addition, the second pmf is specified in such a way that it has again the same expectation, implying  that the mixture has the very same expectation too.

\begin{figure}
	\includegraphics[width=\linewidth]{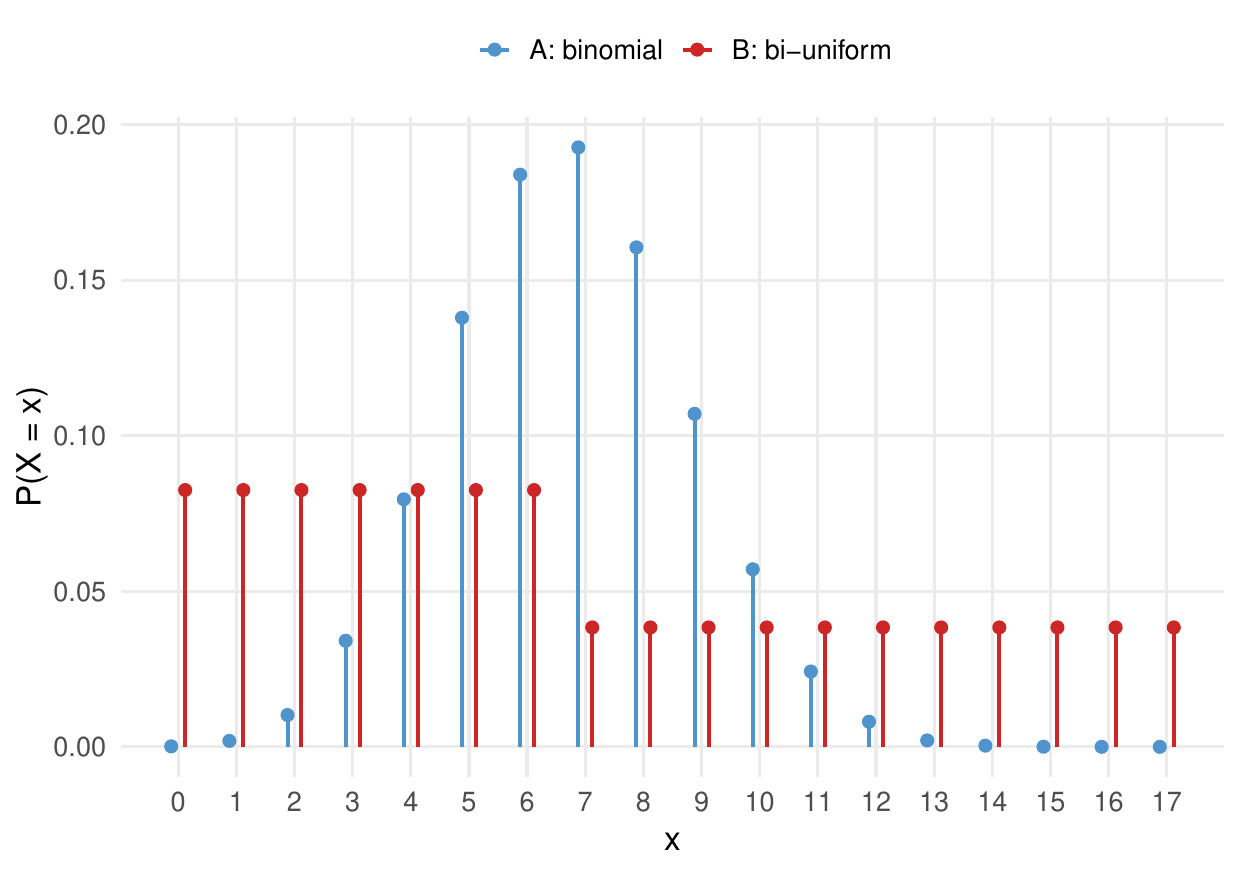}
	\caption{Binomial pmf for the case $n=17$ and $p=0.4$ compared with bi-uniform pmf with $\mu=n p$.} 
	\label{fig1} 
\end{figure}

\section{Bi-Uniform Distribution}\label{sec:biuni}

From the classical collection of discrete distributions a quite simple one defined on the bounded support $M$ is the discrete uniform distribution with pmf
$$
p_{X}(x) = \frac{1}{n+1}
$$
for $x\in M$ and $p_{X}(x) =0$ otherwise. It has expectation
$$
\ex(X) = \sum_{x=0}^{n} x \frac{1}{n+1} = \frac{n(n+1)}{2 (n+1)} = \frac{n}{2}
$$
and variance
$$
\var(X) = \sum_{x=0}^{n} x^2 \frac{1}{n+1} - \frac{n^2}{4} = \frac{n^2 + 2 n}{12}\; .
$$

Assume now that a very similar pmf is sought with the property that its expectation is not necessarily in the center but may take any specified value between $0$ and $n$. A successful approach allows the usage of only two different probability values instead of $1/(n+1)$.
To see this, let $\mu$ be any real number with $0 \leq \mu < n$ and let $\lfloor \mu \rfloor$ be the integer part of $\mu$. Then the definition
$$
p_{X}(x) = \begin{cases}
p_{1} &  \text{for $x\in \{0,1,\ldots, \lfloor \mu \rfloor\}$}\\
p_{2} & \text{for $x\in \{\lfloor \mu \rfloor+1,\ldots, n\}$}\\
0 & \text{otherwise}
\end{cases}
$$
gives a pmf which will be called the bi-uniform distribution with parameter $\mu$ on the support $M$.
The probabilities $p_{1}$ and $p_{2}$ are chosen such that
$$
1 = (\lfloor \mu \rfloor +1) p_{1} + (n-\lfloor \mu \rfloor) p_{2}\; ,
$$
and such that the expectation equals the desired value $\mu$, meaning
\begin{eqnarray*}
\mu & = & p_{1} \sum_{x=0}^{\lfloor \mu \rfloor} x + p_{2} \sum_{x = \lfloor \mu \rfloor+1}^{n} x\\
& = & (p_{1} -p_{2}) \sum_{x=0}^{\lfloor \mu \rfloor} x - p_{2} \sum_{x = 0}^{n} x\\
& = & (p_{1} -p_{2}) \frac{\lfloor \mu \rfloor(\lfloor \mu \rfloor +1)}{2} - p_{2} \frac{n(n+1}{2}\; .
\end{eqnarray*}
For example for $n=17$ and $p=0.4$ the two required equations can be comprised into the system
$$
\begin{bmatrix}
21 & 132\\
7 & 11
\end{bmatrix} \begin{bmatrix}
p_{1}\\
p_{2}
\end{bmatrix} = \begin{bmatrix}34/5 \\ 1
\end{bmatrix}\; .
$$
It admits the solutions $$p_{1} = \frac{26}{315}, \quad p_{2} = \frac{19}{495}$$ also displayed in Figure \ref{fig1}.
For the general case it is seen that explicit solutions are given by
$$
p_{1} = \frac{n+1 -(2\mu-\lfloor \mu \rfloor)}{(\lfloor \mu \rfloor+1)(n+1)}, \quad p_{2} = \frac{2 \mu - \lfloor \mu \rfloor }{(n-\lfloor \mu \rfloor)(n+1)}\; .
$$
These solutions are in fact probabilities lying between 0 and 1.
Obviously, the choice $\mu=n/2$ gives $p_{1} = p_{2} =1/(n+1)$, i.e. the classical discrete uniform distribution on $M$.

The bi-uniform pmf is introduced here not with the intention of proposing a distribution for actual applications, but for providing a tool useful in putting more weight on the tails by weighting it with the binomial pmf itself. From the comparison of the two pmfs displayed in Figure \ref{fig1} for the special case $n=17$, $p=0.4$ and $n p = \mu = 6.8$, it is natural to assume that a mixture can do the job. More explanations are given in  Section \ref{sec:mixture} and an application is discussed in Section \ref{sec:app}.

\subsection{Variance}

Since the parameter $\mu$ is identical to the expectation of the bi-uniform distribution,
the variance is specified by the formula
$\var(X) = \ex(X^2) - \mu^2$. For the bi-uniform on $M$,
\begin{eqnarray*}
\ex(X^2)
 & = & (p_{1} - p_{2}) \sum_{x=0}^{\lfloor \mu \rfloor} x^2 + p _{2} \sum_{x=0}^{n} x^2\\
& = & (p_{1} - p_{2}) \frac{\lfloor \mu \rfloor(\lfloor \mu \rfloor+1)(2\lfloor \mu \rfloor+1)}{6}\\
& & \phantom{(p_{1}} + p_{2}\phantom{)} \frac{n(n+1)(2n+1)}{6}\; .
\end{eqnarray*}
Collecting terms and reducing fractions eventually yields 
$$
v_{2} := \var(X)  = \frac{\mu (2n+1) - 3 \mu^2 + (2\mu -n) \lfloor \mu \rfloor}{3}\; .
$$
If $\mu= n/2$ (not necessarily an integer), $v_{2}$ is identical to the variance of the
classical uniform distribution on $M$.

\subsection{Variance Difference}\label{sec:diff}

\begin{figure}
	\includegraphics[width=\linewidth]{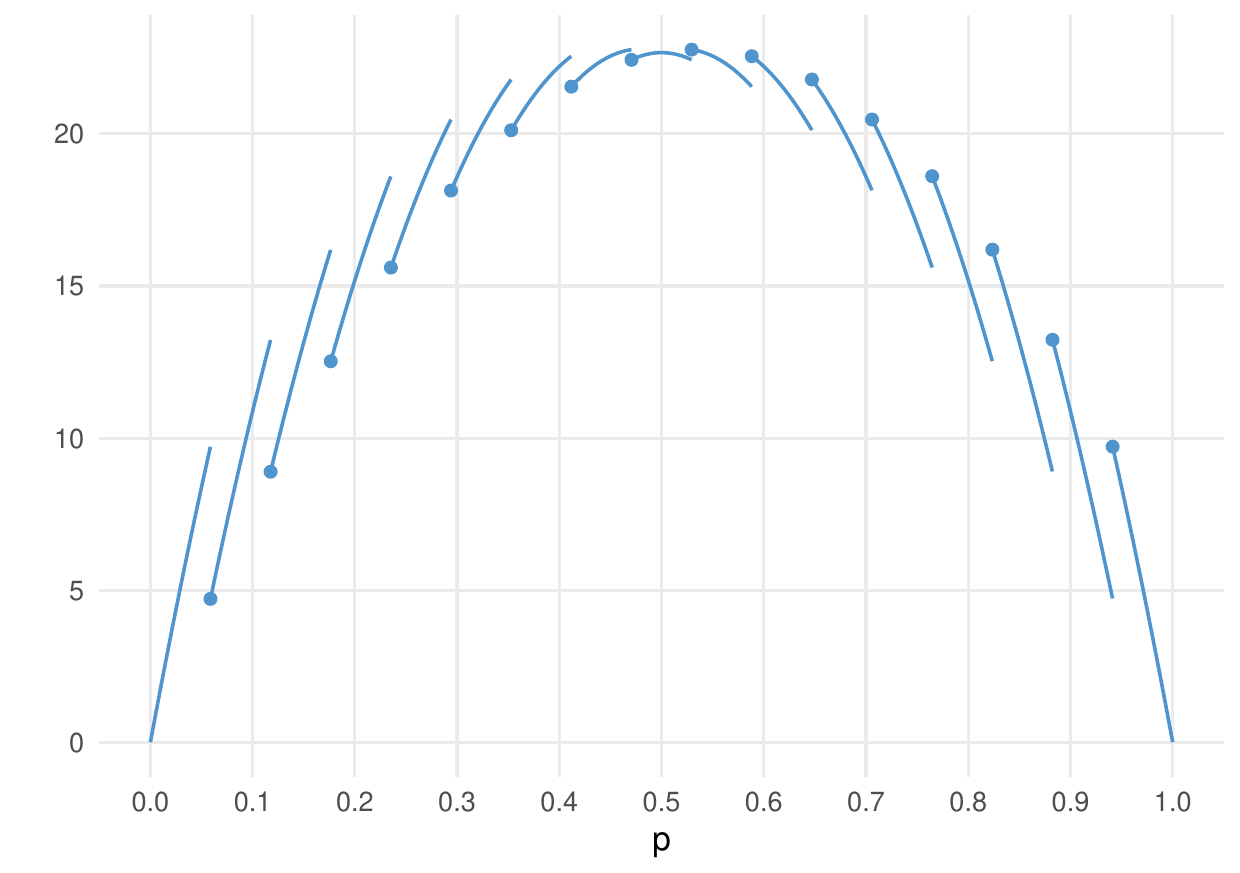}
	\caption{Variance of the bi-uniform pmf minus variance of the binomial pmf, i.e. $d(n=17, p)$ as a function of $0<p<1$.} 
	\label{fig2} 
\end{figure}

If $\mu=n p$ for $1 < n$ and $0< p <1$, the variance difference $$d(n,p) := v_{2} - v_{1}$$
can be shown to be positive. When
$$
\{n p \} = np - \lfloor np\rfloor
$$
denotes the fractional part of $np$ with $0\leq \{np\}<1$, after some rearrangements
$$
d(n,p) = \frac{\eta(n,p) + \{n p \} n(1-2p)}{3},
$$
where
$$
\eta(n,p) = np \left(n(1-p) + 3p -2\right)\; .
$$
In view of $n \geq 2$ ist follows
$$
\eta(n,p) \geq  np \left(2(1-p) + 3p -2\right) = n p^2 > 0 \; .
$$
This implies that if $\{np\} = 0$, or $p \leq 1/2$ then $d(n,p) > 0 $.
If  $\{np\} \not=0$,  then $\lfloor n(1-p) \rfloor = n - \lfloor n p \rfloor -1$ and hence
$\{n(1-p)\} = 1  - \{np\}$. In that case $d(n,p) = d(n, 1-p)$, which is also illustrated by the graph of
$d(n,p)$ for $n=17$ displayed in Figure \ref{fig2}. This also implies that  if $\{np\}\not=0$ and $p>1/2$, then $d(n,p) >0$.

\section{Mixture Distribution}\label{sec:mixture}

Suppose now that $p_{X}^{(1)}(x)$ denotes the pmf of the binomial distribution with parameters $n$ and $p$ while $p_{X}^{(2)}(x)$ stands for the pmf of the bi-uniform distribution on the same support $M=\{0,1, \ldots, n\}$ with parameter $\mu = np$.
If we let $w$ be some weight factor such that $0\leq w \leq 1$, then  the mixture
$$
p_{X} (x) = w \cdot p_{X}^{(1)}(x) + (1-w) \cdot p_{X}^{(2)}(x)
$$
is again a pmf on $M$. Moreover, it has expectation $\ex(X) = np$ and variance $$
\var(X) = w \cdot v_{1} + (1-w) \cdot v_{2} \; .
$$
In general, the variance of a mixture is not the weighted averages of the individual variances, see e.g.
Section 3.7 in  \citet{hogg2019introduction}, but here it holds true because both pmfs have the same expectation.

The shape of the mixture pmf is limited to some form in between the binomial and the bi-uniform. Therefore this mixture is not as flexible as the beta-binomial pmf which may for example also take some U-shape. For applications the mixture will be most appropriate when the empirical frequency distribution in principle resembles the binomial pmf but with more weight in the tails.  In that case a mixture with $w$ close but not exactly equal to $1$ can provide a better fit than the binomial model. This mixture is then a binomial-like distribution with heavier tails as suggested by the title of the paper.

\subsection{Estimation}

The above specified mixture pmf has two parameters $p$ and $w$, which for the fitting process are to be estimated from $m$ data points $x_{1}, \ldots, x_{m}$.

\subsubsection{Method of Moments} The method of moments estimate for the parameter $p$ from the mixture
is the solution
$$
\widehat{p} = \overline{x}/m
$$
to the equation $\ex(X) = \overline{x}$ and thereby the same as under the pure binomial model. Replacing $p$ by $\widehat{p}$
and solving the  equation $\var(X) = s^2$ for $w$ gives
$$
\widehat{w} = \frac{\widehat{v}_{2} - s^2}{\widehat{v}_{2} - \widehat{v}_{1}}\, ,
$$
where $\widehat{v}_{j}$ are the $v_{j}$, $j=1,2$, with $p$ replaced by $\widehat{p}$. As shown in Section \ref{sec:diff} the denominator is strictly positive for $1 < n$ and  $0 < \widehat{p} <1$.

It is readily seen that $0\leq w\leq 1$ if and only if
$$
\widehat{v}_{1}^2 \leq s^2 \leq v_{2}^2\; .
$$

The two inequalities are satisfied when extra-binomial variance is present (when assessed by $\widehat{v}_{1}^2 \leq s^2$, but see Section \ref{sec:over}),
while not being stronger than what can be captured by the bi-uniform distribution (when measured by  $s^2 \leq v_{2}^2$).

In principle it is possible that $s^2 > \widehat{v}_{2}$. For an example let $n=17$ and assume that there are 10 out of $m= 20$ observations equal to 0 and the other 10 equal to 17. Then $\widehat{p}=1/2$ and $\widehat{v}_{2} \approx 26.92$ but $s^2 \approx 76.05$. However, this is an extreme example where the empirical frequencies in no way resemble binomial probabilities.

\begin{table}
	\caption{Frequencies for families with eight children and expected frequencies from two fitted models.}\label{tab:fisher}
	\centering
	\begin{tabular}{rrrr}
		\toprule
		\multicolumn{2}{c}{Counts} &\multicolumn{2}{c}{Expected} \\
		\cmidrule(r){1-2} \cmidrule(r){3-4}
		Sons & Families & Binomial & Mixture  \\
    \midrule
0 &     215 &   165.2 &   231.3\\
1 &    1485 &  1401.7 &  1453.0\\
2 &    5331 &  5202.6 &  5208.4\\
3 &   10649 & 11034.7 & 10970.6\\
4 &   14959 & 14627.6 & 14520.5\\
5 &   11929 & 12409.9 & 12336.9\\
6 &    6678 &  6580.2 &  6577.1\\
7 &    2092 &  1993.8 &  2045.5\\
8 &     342 &   264.3 &   336.8\\
\midrule
Total & 53680 & 53680.0 & 53680.1 \\
		\bottomrule
	\end{tabular}
\end{table}

\subsubsection{Minimum Chi-Square}\label{sec:minchi}

Using the above method of moments estimate $\widehat{p}$ guarantees that when comparing the pure binomial with the mixture fit, both have the same estimated expectation. 

The parameter $w$ may nonetheless alternatively be estimated by the so-called method of minimum chi-square. See e.g. \citet{harris1983use} for a general review and insightful remarks. The method employs the Pearson chi-square statistic given here by
$$
\chi^2 = \sum_{x=0}^{n} \frac{[O(x) - E(x)]^2}{E(x)}\;,
$$
where $O(x)$ denotes the absolute frequency observed for $x\in M$ and
$$
E(x) = m \cdot p_{X}(x)
$$
denotes the estimated expected frequency for $x$ from a pmf $p_{X}(x)$.

Given the estimate $\widehat{p}$ for $p$, one may now choose
$w$ such that $\chi^2$ becomes minimal.
When the goodness of the fit is measured by $\chi^2$ this guarantees an optimal fitting by simultaneously ensuring that the estimated expectation remains unchanged.

The general idea of adapting a standard model by using weighted probabilities in order to achieve a better fit is also portrayed by \citet{harris1983use} who discuss a rather complex weighting procedure for a specific data set by using the truncated geometric distribution as a standard first stage model.

\section{Application}\label{sec:app}

Table \ref{tab:fisher} displays a classical data set discussed as Example 6 in \S 18 by \citet{fisher1958statistical} and also re-investigated by \citet{ishii1960compound} for the application of the beta-binomial distribution. \citeauthor{fisher1958statistical} refers to it as ``Geissler's data on the sex ratio in German families'' \citep[p. 66] {fisher1958statistical}. For $m= 53680$ families with $n=8$ children the number of male offspring is recorded and a binomial model is considered with the parameter $p$ being the probability for a son.

\subsection{Binomial Model}

The usual estimator under the binomial model yields $$\widehat{p} = 0.51468$$ and the estimated variance is $\widehat{v}_{1} = 1.99828$. The sample variance is computed as
$s^2 =  2.06745$, all numbers rounded to 5 digits after the dot. The binomial index of dispersion comes out as $$z
= (m-1) s^2 / \widehat{v}_{1} = 55537.31\, $$
while the $0.95$ quantile of the chi-square distribution with $m-1$ degrees of freedom is 54219.08. Hence, according to Section \ref{sec:over} these values suggest some meaningful extra-binomial variance. \citeauthor{fisher1958statistical} notes that one possible explanation could be the occurrence of multiple births which are known to be more often than not of the same gender. Table \ref{tab:fisher} shows, among other characteristics also discussed by \citeauthor{fisher1958statistical}, that frequencies of $0$ and $8$ are larger than predicted from the binomial model.

\subsubsection{Minimum Chi-Square Estimation}

In principle it is also possible to estimate the parameter $p$ from the pure binomial model by the method of minimum chi-square as introduced in Section \ref{sec:minchi}.  The resulting estimate is not necessarily identical to the method of moments estimate. For the given data, it turns out as $\widetilde{p} = 0.51470$, which is the same as $\widehat{p}$ when both are rounded to four digits after the dot.

\begin{table}
	\caption{Chi-square statistic and corresponding effect size measure from four fitted models.}\label{tab:goodness}
	\centering
	\begin{tabular}{lrr}
		\toprule
		& $\chi^2$ & $w_{\text{ES}}$\\
\midrule
Binomial & 91.87 & 0.0414\\
Mixture  & 43.58 & 0.0285\\
Mixture (mom) & 45.47 & 0.0291\\
Beta-Binomial & 53.31 & 0.0315\\
		\bottomrule
	\end{tabular}
\end{table}

\subsection{Beta-Binomial Model}

As motivation for considering the beta-binomial model \citet{ishii1960compound} refer to the sex-ratio example from the book by  \citeauthor{fisher1958statistical}. The method of moment estimates for the parameters are
 $$
 \widehat{\alpha} = 103.5537, \quad \widehat{\beta} = 97.6475\; .
 $$
 Table \ref{tab:goodness} shows that the $\chi^2$ value from the beta-binomial model is noticeably smaller than for the binomial fit. In Figure \ref{fig5} the individual term contributions to $\chi^2$ are graphically illustrated. The beta-binomial model mostly lacks fit in the center but slightly also for the extreme values $0$ and $8$.

\subsection{Mixture Model}

\begin{figure}
	\includegraphics[width=\linewidth]{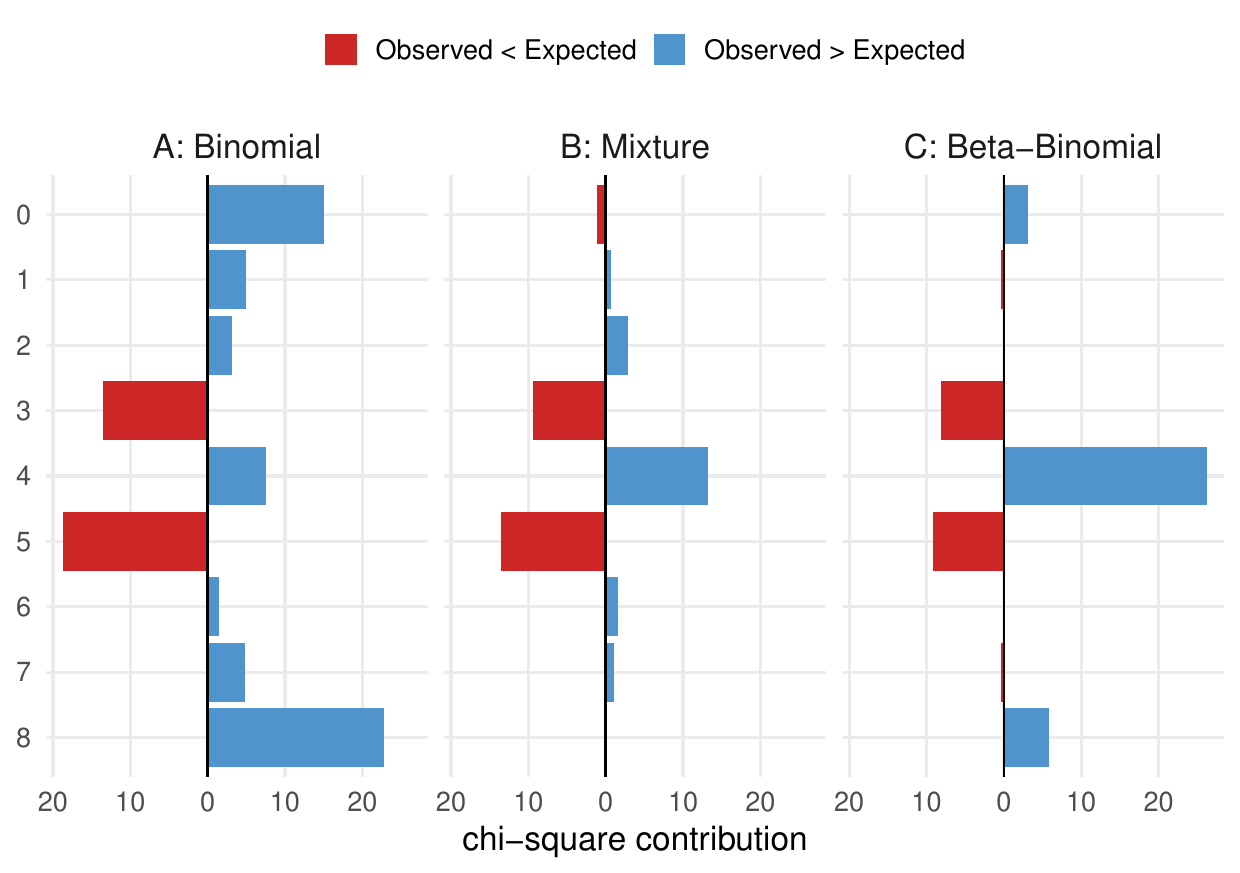}
	\caption{Contributions to the chi-square statistic from three fitted models.}
	\label{fig5} 
\end{figure}

In Tables \ref{tab:fisher} \& \ref{tab:goodness} and in Figure \ref{fig5}, model ``Mixture'' refers to the mixture of binomial and bi-uniform with mixing weight factor $\widehat{w} = 0.98802$  from the minimum chi-square method. For simplicity, $\widehat{w}$ had been determined as the minimizing value of $\chi^2$ over a sequence of possible values with a distance of 0.00001 to each other. Table \ref{tab:goodness} shows that this model yields the smallest $\chi^2$ value among the considered models. In Table \ref{tab:fisher} the expected frequencies can be compared with the observ counts and with the expected frequencies from the binomial model. From Figure \ref{fig5} it is seen that the contributions to $\chi^2$ are quite small in the tails but still somewhat larger in the center.

The weight factor determined by the method of moments is $\widetilde{w} = 0.98526$. As it seen from Table \ref{tab:goodness} the corresponding model ``Mixture (mom)'' yields a slightly larger $\chi^2$ value.

\subsection{Goodness-of-fit}

The $\chi^2$ statistic from the previous section is often computed to carry out a good\-ness-\-of-\-fit test. The null hypothesis that the specified distribution generates the data is rejected at a specified level $\gamma$ when $\chi^2$ is greater than the
$(1-\gamma)$ quantile of the chi-square distribution with $n-k$ degrees of freedom. Here $k$ is the number of parameters to be estimated, and $n$ is minus one the number of categories $0,1,\ldots , n$.

The degrees of freedom for the binomial model are therefore $8-1$, while they are $8-2$ for the beta-binomial  and the mixture model. For $\gamma=0.05$ the (rounded) critical values are 14.07 for 7 degrees of freedom and 12.59 for 6 degrees of freedom. Hence, the null hypothesis is rejected for all distributional model fits in question.

\subsubsection{Effect Size}

On the other hand, it is well known that for large sample sizes statistical hypothesis tests yield significant results even for very small deviances from the null hypothesis. To counteract this, effect size (ES) measures are often used as additional tools to help assessing whether significant results are in fact also relevant. These measures had been popularized by \citet{cohen1988statistical} who discusses
$$
w_{\text{ES}} = \sqrt{\chi^2 /m}
$$
(known as Cohen's $w$) as a possible ES measure for the $\chi^2$ goodness-of-fit test. Values smaller than 0.1 are suggested to indicate a small to less than small effect, although \citeauthor{cohen1988statistical} cautions the reader against uncritical use of $w_{\text{ES}}$ and states: ``The investigator is best advised to use the conventional definitions as a general frame of reference for ES and not to take them too literally.'' \citep[p. 224]{cohen1988statistical}. Effect size measures for the above models are given in Table \ref{tab:goodness}.

\subsection{Conclusion}

The above considerations show that the mixture model provides a good fit to the given data.  It seems that this model can work best for instances where a weight factor close to 1 is appropriate.  
There may of course be various other data circumstances under which the application of the more flexible beta-binomial distribution will be more suitable. 

\bibliographystyle{agsm}
\bibliography{biunifmix}

\end{document}